\documentclass[12pt]{article}

\usepackage{ifpdf}
\usepackage{hyperref}
\usepackage{pstricks}
\usepackage{color}
\usepackage{amssymb,amsmath,amsfonts,bm,bbm}
\usepackage{afterpage}
\usepackage{longtable}
\usepackage{cite}
\usepackage{latexsym, mathrsfs}
\usepackage{graphicx,graphics,epsf,epsfig,epstopdf,subfigure,psfrag}
\usepackage{url}
\usepackage{paralist}
\usepackage{bbold}

\ifpdf

\newcommand{\be}{\begin{equation}}
\newcommand{\ee}{\end{equation}}
\newcommand{\bea}{\begin{eqnarray}}
\newcommand{\eea}{\end{eqnarray}}
\newcommand{\bec}{\begin{center}}
\newcommand{\eec}{\end{center}}
\newcommand{\nn}{\nonumber}

\begin{document}

\begin{flushright}{BARI-TH/2015-697} \end{flushright}

\thispagestyle{empty}
\bigskip

\begin{center}
{\LARGE\bf Holographic Oddballs}\\[1.0 cm]
{\bf  L.~Bellantuono$^{a,b}$, P.~Colangelo$^{b}$,  F.~Giannuzzi$^{a}$\\[0.5 cm] }
{\small
$^a$Dipartimento di Fisica, Universit\`a  di Bari,via Orabona 4, I-70126 Bari, Italy\\
$^b$INFN, Sezione di Bari, via Orabona 4, I-70126 Bari, Italy\\}
\end{center}

\bigskip

\abstract{The spectrum  of  the glueball with $J^{PC}=0^{--}$   is computed using different bottom-up holographic models of QCD. The results indicate a lowest-lying  state  lighter than in the determination by other methods, with  mass $m \simeq 2.8$ GeV.  The in-medium properties of this gluonium are investigated, and stability   against thermal and density effects is compared to other hadronic systems.  Production and decay modes are identified,  useful for   searching   the $J^{PC}=0^{--}$ glueball. }

\vspace*{1.0cm}

 Glueball and nonstandard multi-quark/gluon states. 

\vspace*{1cm}


\section{Introduction}
The existence of  bound states  of gluons (the so-called  "gluonia" or "glueballs"), with a rich spectroscopy and  a complex phenomenology,  is one of the early  predictions
resulting  from the non-abelian nature of strong interactions described by QCD \cite{Jaffe:1975fd}.
However,  after about 40 years of experimental efforts, evidence of no one of such gluonic states has been unambiguously established \cite{Ochs:2013gi}.
The  glueball with $J^{PC}=0^{++}$, expected to be the lightest one, shares the vacuum quantum numbers with  $\bar q q$ states; as a consequence,  hadrons can result from the mixing between quark-antiquark  and gluonium components. The scalar  isoscalar  mesons $f_0(1370)$, $f_0(1500)$ and $f_0(1710)$  are candidates for such  light mixed states,  with   uncertain mixing angles  inferred from  the production processes and decay modes \cite{Cheng:2015iaa}.
The $J^{PC}=2^{++}$ glueball,  predicted to   be heavier than the scalar one, can  mix with conventional $\bar q q$ $P$-wave configurations, therefore also the identification of  the  tensor glueball  is difficult. A signature for the experimental investigations
is that  gluonia   overpopulate the meson multiplets with fixed quantum numbers. Moreover,   since glueballs are $SU(3)_F$ singlets, they are expected to equally couple  to $u, d$ and $s$ quarks
with clear predictions of the decay fractions  in pions, etas and kaons.  However, 
the chiral behaviour of such couplings needs  also to be taken into account:   it  can induce  deviations  from the flavor-symmetric condition due to a quark mass dependence,  as discussed for the  $J^{PC}=0^{++}$ state in \cite{Chanowitz:2005du}, and  this makes  the glueball  identification further involved.

Remarkably, there is the possibility of having gluonia with combinations of $J^{PC}$ not allowed in the quark model; therefore,  searching for states with such "exotic"  quantum numbers is  a strategy to look  for 
 gluonic resonances. An interesting case is  $J^{PC}=0^{--}$: glueballs which these quantum numbers, as  for all states with negative charge conjugation, must be composed by an odd number of constituent gluons,  which justifies the 
 name of  "oddballs". Although $C$-odd gluonia are expected to be heavier than the scalar glueball,   they  are  within the reach of the  present-day experimental  facilities.

Little theoretical information is available about gluonic resonances with $J^{PC}=0^{--}$. In the old flux-tube model, Isgur and Paton predicted  $m_{0^{--}}=2.79$ GeV, with the  mass ordering in the gluonium spectrum: 
$m_{0^{++}}<m_{0^{--}}\sim m_{2^{++}}$  \cite{Isgur:1984bm}. Lattice QCD simulations have reported a large value  for the mass: $m_{0^{--}}=5.166$ GeV, with an estimated uncertainty of 1 GeV;
in the same  calculation,   the mass of the  lightest glueball is $m_{0^{++}}=1.795$ GeV, with  $3.3 \%$ uncertainty 
 \cite{Gregory:2012hu}. Two stable $0^{--}$ oddballs have been obtained using QCD sum rules,   with mass $m_{0^{--}}=3.81\pm0.12$ GeV and  $m_{0^{--}}=4.33\pm0.13$ GeV, respectively    \cite{Qiao:2014vva}. 
In all cases, the width and the hadronic couplings  are unknown.
On the experimental side,  analyses of the $D^0 \to \pi^+ \pi^- \pi^0$ decay mode indicate that the  final state is nearly completely dominated  by  a $I=0$,  $J^{PC}=0^{--}$ configuration \cite{Malde:2015mha}; 
however, the confirmation of   the contribution  of a resonance close to $D^0$  (the $D^0$ mass is $1864.84$ MeV), together with the interpretation of this puzzling result,   requires a further scrutiny.

An interesting  method for computing various hadronic properties is inspired by  the AdS/CFT (Anti-de Sitter/conformal field theory) correspondence   \cite{Maldacena:1997re,Witten:1998qj,Gubser:1998bc}. 
In this approach, at large 't Hooft coupling and in the limit of  large number of colors $N_c$, the correlation functions of gauge-invariant 
operators in a $4D$ gauge field theory are derived by a classical gravity theory in a higher dimensional space.\footnote{An  overview can be found in \cite{Ammon:2015wua}.}
 The calculation of the mass of the scalar  $J^{PC}=0^{++}$ and  tensor $J^{PC}=2^{++}$ glueballs is  one of the first applications of the method
\cite{Gross:1998gk,Csaki:1998qr,Hashimoto:1998if,Constable:1999gb,Brower:2000rp},  starting from the top-down construction  based on a type II-A supergravity with supersymmetry and  conformal invariance broken,  and  a Yang-Mills theory at large $N_c$  as a dual  \cite{Witten:1998qj}.
 Including matter fields, the analysis has been extended to the glueball  hadronic couplings  \cite{Hashimoto:2007ze,Brunner:2015oqa}.   The mass spectrum of the $C-$odd $J^{PC}=1^{+-}$ and $1^{--}$ glueballs has  been computed  in the top-down construction, too \cite{Brunner:2015oqa}.

The gauge/gravity duality method has also been applied  in a more phenomenological procedure which  attempts to formulate,  through  a  bottom-up construction, higher dimensional models  able to reproduce  the largest number of QCD properties.
Here, we  follow this bottom-up approach which has been used  to describe conventional $\bar q q$  \cite{Erlich:2005qh,DaRold:2005zs,Karch:2006pv,Colangelo:2008us,Brodsky:2014yha} and  
hybrid mesons \cite{Bellantuono:2014lra}, as well as  glueballs \cite{BoschiFilho:2002ta,Colangelo:2007pt,Forkel:2007ru,Colangelo:2007if}. We focus on the lightest oddball with the aim of determining  properties like the mass spectrum. 

In Sect.~\ref{sec2} we compute the spectrum of the $J^{PC}=0^{--}$ gluonium in three different holographic models of QCD,   identifying  robust predictions,  discussing  uncertainties and  proposing  improvements. 
Other interesting aspects to investigate concern the properties of the  glueball in a thermalized and dense hadronic medium. Pointing out the differences with respect to ordinary $\bar q q$ mesons provides us with a better understanding  of this gluonium state, namely about its stability features. The  calculation is affordable in the holographic framework, and is carried out 
in Sect.~\ref{sec3}.
From the experimental viewpoint, it is important to identify the main production processes and the decay modes useful for the searches of the $J^{PC}=0^{--}$ glueball: this is done in Sect.~\ref{sec4}, before  presenting the conclusions of our study.

\section{Oddballs in bottom-up AdS/QCD }\label{sec2}

 In gauge/gravity inspired models of QCD, the starting point is the association of gauge invariant boundary theory operators to fields defined on an $AdS_5$ manifold, with modifications needed to describe confinement.
In QCD,  composite gauge-invariant local operators with quantum numbers $J^{PC}=0^{--}$,  involving only gluon fields,  can be written 
in terms of  the gluon field strength $G^a_{\mu \nu}(x)$ and  the dual $\widetilde G^a_{\mu \nu}(x)=\frac{1}{2} \epsilon_{\mu \nu \rho \sigma}G^a_{\rho \sigma}(x)$:
\bea
J^A(x)&=&g_s^3 d_{abc} [\eta^t_{\alpha \beta} \widetilde G^a_{\mu \nu}(x)] [\partial_{\alpha}  \partial_{\beta} G^b_{\nu \rho}(x)] [G^c_{\rho \mu}(x)] \nn \\
J^B(x)&=&g_s^3 d_{abc} [\eta^t_{\alpha \beta}  G^a_{\mu \nu}(x)] [\partial_{\alpha}  \partial_{\beta} \widetilde G^b_{\nu \rho}(x)] [G^c_{\rho \mu}(x)] \nn \\
J^C(x)&=&g_s^3 d_{abc} [\eta^t_{\alpha \beta} G^a_{\mu \nu}(x)] [\partial_{\alpha}  \partial_{\beta} G^b_{\nu \rho}(x)] [\widetilde G^c_{\rho \mu}(x)] \label{odd-op} \\
J^D(x)&=&g_s^3 d_{abc} [\eta^t_{\alpha \beta} \widetilde G^a_{\mu \nu}(x)] [\partial_{\alpha}  \partial_{\beta} \widetilde G^b_{\nu \rho}(x)] [\widetilde G^c_{\rho \mu}(x)]  \,\,\, ,\nn 
\eea
with $a,b,c$  color indices, $d_{abc}$ the symmetric $SU(3)_c$ structure constants,  and  $g_s$   the strong coupling constant.
The transverse  $\eta^t_{\alpha \beta}$ metric is defined as $\eta^t_{\alpha \beta}= \eta_{\alpha \beta} - \frac{\partial_\alpha \partial_\beta}{\partial^2}$, with   $\alpha,\beta$ (as well as $\mu,\nu$) $4D$ Lorentz indices,  and $\eta_{\alpha \beta}$ the Minkowski metric tensor  \cite{Qiao:2014vva}.

We focus on only one operator in \eqref{odd-op}, generically denoted as $J_0(x)$, which has conformal dimension $\Delta=8$, and   associate to $J_0(x)$ a  dual field in $AdS_5$, 
$O_0(x,z)$,  with mass obtained by the relation $M_5^2 R^2=\Delta (\Delta-4)$ \cite{Witten:1998qj,Gubser:1998bc}. $R$ is  the $AdS_5$ radius; we set $R=1$. 

We choose Poincar\`e coordinates $x_M=(x_\mu,z)=(x_0,\vec x, z)$ for the $AdS_5$ space, with line element  
\be
ds^2=g^{MN}dx_M dx_N= \frac{1}{z^2}\left( dx_0^2-d\vec x^2-dz^2 \right) \label{AdS-geom}
\ee
$(0 < z)$, and define an action for the field $O_0(x,z)$.  
To account for  confinement in  QCD, in the definition of the action the  breaking of  conformal invariance must be implemented: this can be done in different ways, some of which are used in the following.

\subsection{Hard-wall model (HW)}
A simple way of modeling confinement in the holographic setup is  by considering a slice of the AdS$_5$ space, with  a sharp cutoff  at a finite distance $z_m$ along the extra dimension 
\cite{Erlich:2005qh,DaRold:2005zs}.  The $5D$ action for  $O_0(x,z)$  can then be written as
\be
S_{(HW)}=\frac{1}{k} \int d^5 x  \sqrt g \left[ g^{MN} \partial_M O_0\,  \partial_N O_0 -M_5^2 O_0^2 \right] \,\, ,  \label{action-HW}
\ee
with $\epsilon \leq z \leq z_m$ and $g=|det (g_{MN}) |$.  The constant  $k$  makes  the action dimensionless, and its derivation is presented in the next section. The value of $1/z_m$ sets the hadronic scale in the model, so that the dimensionful  quantities are given in terms of this parameter.

\begin{figure}[t]
\bec
\includegraphics[width = 0.5\textwidth]{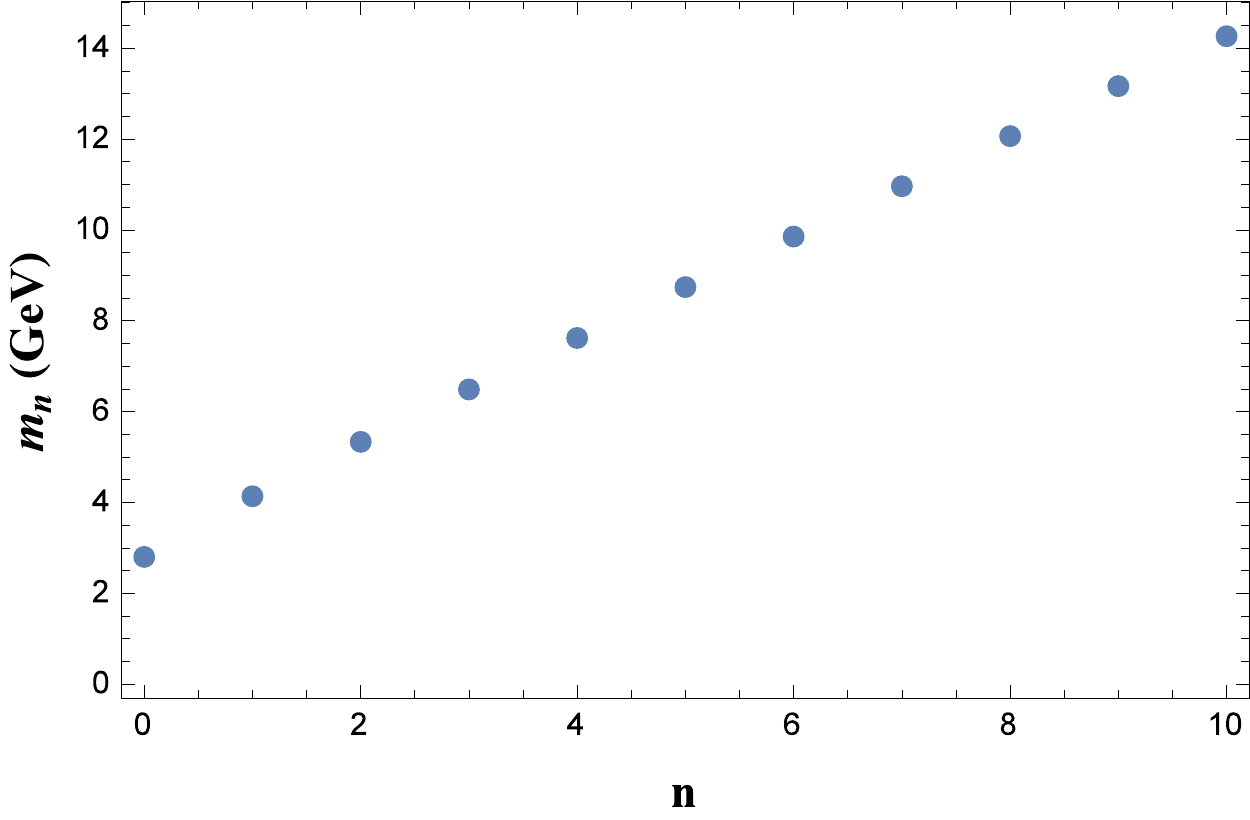}
\caption{\small  Spectrum of the $J^{PC}=0^{--}$  gluonium by the HW model,  with   $1/z_m=346$ MeV. }\label{hw-spettro}
\eec
\end{figure}

The  equation of motion for the Fourier image $\widetilde O_0(p,z)$ of $O_0(x,z)$,
\be
-\frac{p^2}{z^3} \widetilde O_0(p,z) -\partial_z \left[ \frac{1}{z^3}  \partial _z \widetilde O_0(p,z) \right ] + \frac{M_5^2 }{z^5} \widetilde O_0(p,z) =0 \,\,\, , \label{hw-eqm}
\ee
together with the Dirichelet boundary condition  for  $\widetilde O_0(p,z)$ at    the ultraviolet  $z=\epsilon$ brane,  $\widetilde O_0(p,\epsilon)=0$ (with $\epsilon \to 0^+$), and  the Neumann boundary condition  at the infrared  $z=z_m$ brane, $\partial_z \widetilde O_0(p,z_m)=0$,
allows  to compute  the spectrum. The resulting lightest masses,  setting  $1/z_m=346$ MeV, are collected in  Fig.~\ref{hw-spettro}. The  masses of the first two states are
$m_0=2.80$ GeV and   $m_1=4.14$ GeV, respectively.  Such values reduce to $m_0=2.61$ GeV and $m_1=3.86$ GeV if the hadronic scale is  $1/z_m=323$ MeV; both values of $1/z_m$  are used in   phenomenological analyses \cite{Erlich:2005qh}.  A linear dependence is obtained for  $m_n$ vs the radial (in the extra dimension) quantum number $n$,  a feature shared by other hadrons  described by this model.

\subsection{Soft-wall model (SW)}
A different way of  breaking the conformal invariance and modeling  confinement in QCD  consists in including a  background field 
  $\phi(z)$ in the $5D$ action for  $O_0(x,z)$   \cite{Karch:2006pv}. In particular, the profile  $\phi(z)=c^2 z^2$,  which involves a  parameter $c$  fixing the hadronic scale, allows to obtain linear Regge trajectories  and  has been employed to study aspects of  the  light hadron  phenomenology  \cite{Karch:2006pv,Colangelo:2008us}.  The action for the field $O_0(x,z)$ is
\be
S_{(SW)}=\frac{1}{k} \int d^5 x \sqrt g  e^{-\phi(z)} \left[ g^{MN} \partial_M  O_0 \,\partial_N O_0 -M_5^2 O_0^2 \right] \,\,\, , \label{action-SW}
\ee
with $\epsilon<z<\infty$.  
The normalizable solutions of the equation of motion for  $\widetilde O_0(p,z)$,
\be
\frac{e^{-\phi(z)}}{z^3} (-p^2) \widetilde O_0(p,z) -\partial_z \left[ \frac{e^{-\phi(z)}}{z^3}  \partial _z \widetilde O_0(p,z) \right ] + \frac{M_5^2 e^{-\phi(z)}}{z^5} \widetilde O_0(p,z) =0 \,\,\, , \label{sw-eqm}
\ee
 correspond to the  Regge-like  mass spectrum 
\be
m_n^2= 4 c^2 (n+4) \,\,\, . \label{sw-spec}
\ee
\begin{table}
\bec
\begin{tabular}{c  c c c c c}
\hline
$J^{PC}$&$1^{--}$  $(q \bar q)$ \cite{Karch:2006pv} \,&$0^{++}$ $(q \bar q)$ \cite{Colangelo:2008us}&$0^{++}$ (glueball) \cite{Colangelo:2007pt} &$1^{-+}$ \cite{Bellantuono:2014lra}&$0^{--}$ \\
\hline   
 $m^2_n$& $4 c^2 (n+1)$&$4 c^2 (n+3/2)$&$4 c^2  (n+2)$&$4 c^2 (n+2)$ & $4 c^2  (n+4)$\\
\hline
  \end{tabular}
\caption{Mass spectrum of $q \bar q$ vector and scalar states, $0^{++}$  glueball,  $\bar q G q$ $1^{-+}$ hybrid, and $0^{--}$ gluonium, obtained  in the SW model.}\label{tab:spectrum}
\eec
\end{table}
The gluonium turns out to be  heavier than conventional $\bar q q$ mesons,  than the $J^{P}=1^{-+}$ hybrid  and the  $0^{++}$	glueball,  as
one  infers from  the spectral relations  in Table \ref{tab:spectrum}.  However, the mass difference  with respect to   ordinary hadrons is smaller than obtained by  different calculations. 
 Setting $c=m_\rho/2=388$ MeV  from the $\rho$ meson mass  \cite{Karch:2006pv},  we get $m_0=1.55$ GeV and  $m_1=1.74$ GeV, while the value $c=474$ MeV  reproducing the $\rho-$meson Regge trajectory corresponds to $m_0=1.90$ GeV and $m_1=2.12$ GeV.   Hence, the SW model indicates  a  light oddball, as    for  other mass predictions in the same framework.

The two-point correlation function of  $J_0(x)$ in QCD
\be
\Pi(p^2)= i \int d^4 x \, e^{i p x} \langle 0| T[J_0(x) J^\dagger_0(0)]  | 0 \rangle  \,\,\,  \label{qcd-corrfun}
\ee
  can be computed  using the AdS/CFT dictionary identifying $J_0(x)$ as the source of $O_0(x,z)$.
We define the (4D Fourier transformed) bulk-to-boundary propagator $\widetilde K(p,z)$ of the glueball field using the equation: 
\be \widetilde O_0 (p,z)=\widetilde K(p,z) \tilde J_0(p) \,\, , 
\ee
with  $\widetilde O_0(p,z)$ and $\widetilde J_0(p)$ the 4D Fourier transformed bulk field and source, respectively, and  differentiate twice  the on-shell action \eqref{action-SW} with respect to $ J_0$. The  two-point correlation function, obtained for  $ J_0 \to 0$, 
\be
\Pi(p^2)= \frac{2}{k} \left[ \frac{e^{- \phi(z)}}{z^3}  \widetilde K(p^2,z) \partial_z \widetilde K (p^2,z) \right]_{z \to 0} \,\,\, , \label{AdS-corr}
\ee
can be written as
 \be
\Pi(p^2)=\sum_n \frac{R_n}{p^2-{m_n}^2} \,\,  \label{corrf}
\ee
 with  the residues 
\be
R_n=-\frac{4}{15} \frac{(n+6)!}{6! \, n!} \frac{c^{14}}{k} \,\,\, .\label{residues}
\ee
Comparing the $p^2 \to-\infty$ asymptotic behaviour of $\Pi(p^2)$  in QCD   \cite{Qiao:2014vva} 
\be
\Pi^{QCD}(p^2)= \frac{487 \alpha_s^3}{143 \, 2^6 \, 3^3 \, \pi} \left( -p^2 \right)^6 \log \left( \frac{-p^2 }{\Lambda^2}\right) \label{as-QCD}
\ee
with the  expression \eqref{AdS-corr} in the same limit,  we  fix  $k$ (for $N_c=3$):
\be
k=- \frac{143 \, \pi}{487 \, 2^{10} \, 5^2 \, \alpha_s^3} \,\, . \label{k}
\ee
The same expression holds in the HW model.

\subsection{ Einstein-dilaton model (ED)}
In both the HW and SW models the implementation of the confinement mechanism, with an $AdS_5$ background geometry,  is an input assumption defining  each model.  More elaborated approaches include dynamical fields.
A dynamical holographic model of QCD has been formulated in \cite{Li:2011hp,Li:2013oda}, with a scalar dilaton field $\Phi(z)$   in the bulk  and the 5D gravitation-dilaton coupled action  
analysed. The resulting geometry takes the form:
\be
ds_{(ED)}^2=\frac{e^{2 A_s(z)-\frac{4}{3} \Phi(z)}}{z^2} \left[dx_0^2-d\vec x^2 -dz^2  \right] \,\,\, .  \label{metric-ED}
\ee 
The function $A_s(z)$ introduces a quadratic correction in the warp factor distorting the $AdS_5$ metric,  and is chosen with the expression
 $A_s(z)= \bar c \, {\bar k}^2 \, z^2$.  The profile of   the dilaton   is obtained solving the Einstein equations for the metric-dilaton system, and   with the chosen ansatz for $A_s(z)$, $\Phi(z)$ reads
\be
\Phi(z)=\frac{3}{4} \bar c \, {\bar k}^2 \, z^2 \left(1+\,_2F_2\left(1,1;2,\frac{5}{2};2 \bar c \, {\bar k}^2 \, z^2\right) \right) \, \label{dilaton-ED}
\ee
in terms of the generalized hypergeometric function $_2F_2$.
In the model, the dimensionful parameter $\bar k$ is set to   $\bar k=0.43$ GeV. 
 $\bar c$  is  $\bar c=\pm 1$, and both cases   reproduce,  at finite temperature, QCD bulk thermodynamical properties such as the energy density and the speed of sound.
On the other hand, the analysis of the thermodynamical properties of loop operators favours the positive sign;  hence,   we set  $\bar c=+1$. 

The equation of motion  for the oddball field $O_0(x,z)$ with the   metric \eqref{metric-ED}-\eqref{dilaton-ED}  provides for the two lightest  states  the mass $m_0=2.82$ GeV and $m_1=4.07$ GeV, respectively, 
 close to the result of the HW model.


 The  outcome of the analyses in the three models  is that the mass of the lowest-lying $0^{--}$ state  is sensibly lighter than obtained  in  \cite{Gregory:2012hu,Qiao:2014vva},
with results spanning the range $1.55-2.82$ GeV.  The upper value, obtained in the HW and ED models,  is close to the prediction  of the flux-tube model.  For the first excited state, the predicted mass is in the range
$1.74-4.07$ GeV, with again the lighter value given by the SW model. 
The indication in favour of a  light oddball is a  surprising result with interesting phenomenological implications. Indeed,   one can look for this state in the same class of processes  investigated for searching the $0^{++}$ gluonium, namely radiative quarkonium decays  including charmonium. The second consequence is that  there is enough phase-space for $0^{--}$ decays  with a quite  clear experimental signature.
We discuss both the issues in Sect.\ref{sec4}.

\section{Oddballs in medium}\label{sec3}
 Before  addressing the phenomenology of the lightest oddball, 
it is interesting to use the same holographic machinery to investigate other aspects of this gluonic state, namely its
features  in a thermalized and dense hadronic medium. The aim is to make a comparison with the conventional  light vector meson and with the scalar glueball, inferring 
information on the stability properties  of the oddball against thermal and density fluctuations. 
In the holographic approach, the inclusion of matter effects  is affordable using  appropriate bulk geometries \cite{Ammon:2015wua}. 
For definiteness, we consider the  SW model for which many  results  concerning  other hadrons are available for the comparison \cite{Colangelo:2009ra,Colangelo:2012jy}. 

To investigate  in-medium effects  on the oddball spectrum,  we use the $5D$ Reissner-Nordstr\"om AdS metric (AdS/RN).  The issue of  which phase of QCD is described by this bulk geometry is deferred
to the end of this Section; for the time being, we  consider the possibility of describing a stable or a metastable phase \cite{Herzog:2006ra}.

The AdS/RN geometry is defined by the line element
\be
ds^2_{ (RN)}=g^{MN}_{ (RN)}dx_M dx_N= \frac{1}{z^2}\left( f(z) dx_0^2-d\vec x^2- \frac{dz^2}{f(z)} \right) \,\,\,  \label{RN-geom}
\ee
with the function $f(z)$ given by
\be
f(z)=1- \left( \frac{1}{z_h^4} +q^2 z_h^2 \right) z^4+q^2 z^6  . \label{RN-f}
\ee
%
At $q=0$ the geometry  \eqref{RN-geom},\eqref{RN-f} reduces to the AdS/black-hole metric.
$z_h$ is the position of the outer horizon of the black-hole,  the lowest value of the coordinate $z$  satisfying the condition $f(z_h)=0$. 
Defining $Q=q z_h^3$ and imposing the condition  $0\leqslant Q\leqslant \sqrt{2}$, the  black-hole temperature is 
\begin{equation}
 T=\frac{1}{4\pi}\left| \frac{df}{dz}\right|_{z=z_h}=\frac{1}{\pi z_h} \left(1-\frac{Q^2}{2} \right) \,\,\, .
\end{equation}
In \eqref{RN-f}  $q$ is the charge of the black-hole, which can be holographically related to the quark chemical potential $\mu$.  In
the QCD generating functional,  $\mu$  multiplies the quark number operator $O_q(x)=q^\dagger(x) q(x)$. 
Invoking   the gauge/gravity  correspondence, the  coefficient $\mu$ can be considered as the source of the bulk field associated to $O_q(x)$,   the time component of a $U(1)$ gauge field $A_M(x,z)$. 
The AdS/RN metric results from the gravitational interaction of this $U(1)$ field. Within the SW model, we make use of the  AdS/RN geometry together with the background dilaton characterizing the model. 

To fulfil rotational invariance, the $U(1)$ field  $A_M$ has components
$A_i=0$ for $i=1,2,3,z$ , while the component  $A_0$ has the expansion, for $z \to 0$, 
\begin{equation}
 A_0(z)=\mu-\kappa\, q\, z^2  \,. \label{A0}
\end{equation}
The condition that $A_0$ vanishes at the horizon, $A_0(z_h)=0$, provides a linear  relation between $\mu$ and $q$ (or $\mu$ and $Q$),
\begin{equation}
\mu=\kappa \frac{Q}{z_h}\, \label{mucond}
\end{equation}
in terms of a dimensionless parameter $\kappa$ that   can be determined from various  observables \cite{Lee:2009bya}.  In the following we set $\kappa=1$, giving the quark chemical potential up to a numerical factor.
We also set the  dilaton parameter  $c=1$  and the dimensionful quantities  in units of such a scale.

The equation  for the bulk-to-boundary propagator $\widetilde K(p,z)$,  obtained from the action \eqref{action-SW} with AdS/RN background geometry,   reads:
\begin{equation}\label{eomglue}
\partial_z \left[ \frac{e^{- \phi(z)}}{z^3} f(z)  \partial_z \widetilde K(p,z)\right ] +\frac{e^{- \phi(z)}}{z^3} \left[ \frac{p_0^2}{f(z)} - \vec p^{\,2}\right] \widetilde K(p,z)-\frac{M_5^2 e^{- \phi(z)}}{z^5}\widetilde K(p,z) = 0 \,\,\, .
\end{equation}
In the frame with $\vec p=0$,  defining  $\omega^2=p_0^2$ and using the variable $u=z/z_h$,  we compute
the solution of \eqref{eomglue} with the boundary conditions
\bea
\widetilde K(\omega^2,u)&\sim& \frac{1}{z_h^4 u^4} \,\,\,\quad \quad \hspace*{1.7cm}   (u \to 0) \nn \\
 \widetilde K(\omega^2,u)&\sim& (1-u)^{-i \frac{\sqrt{\omega^2 z_h^2}}{4 -2 Q^2}}\,\,\, \quad \quad    (u \to 1) .  \label{infalling}
 \eea
The latter condition selects the {\it in-falling} solution  near the  horizon. 
\begin{figure}[t!]
\bec
\includegraphics[width = 0.6\textwidth]{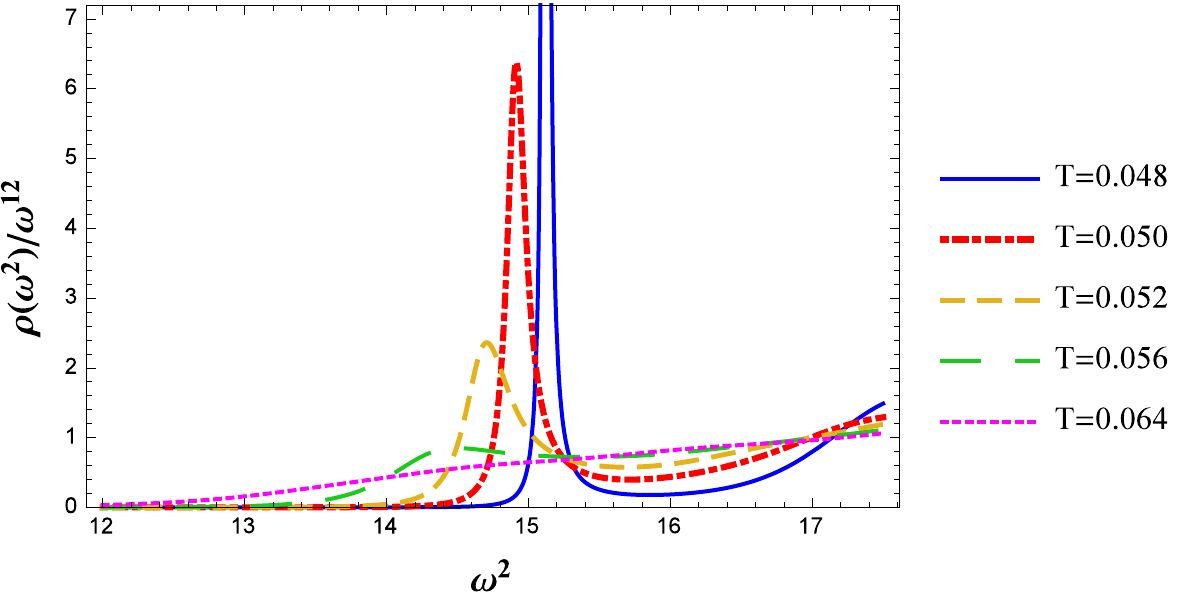}
\caption{\small  Spectral  function for the  $J^{PC}=0^{--}$   oddball in the SW model with the metric  (\ref{RN-geom})-(\ref{RN-f}),  at $\mu=0$ and  for several values of $T$. In the  plotted function $\rho(\omega^2)/(\omega^2)^6$  the constant  
$10^{-8} k/(2 z_h^8)$ has been factorized out. 
The dimensionful quantities are in units of the  scale  $c$. }\label{rho-t}
\eec
\end{figure}
Hence,   the retarded Green's function is  worked out \cite{Son:2002sd,Policastro:2002se,Teaney:2006nc},
\begin{equation}\label{2pfunction}
\Pi^R(\omega^2)= \frac{2}{k} \left[ \frac{e^{- \phi(z)}}{z^3} f(z) \widetilde K(\omega^2,z) \partial_z \widetilde K (\omega^2,z) \right]_{z \to 0} \,\,\, ,
\end{equation}
and  the spectral function 
$ \rho(\omega^2)=\Im \left(\Pi^R(\omega^2)\right)$
is determined   in ranges of  temperature and chemical potential.

At $\mu=0$,  the result in the range of $\omega^2$ corresponding to the lightest resonance is depicted in Fig.~\ref{rho-t}, where we plot    $\rho(\omega^2)/(\omega^2)^6$ (factorizing the constant  $10^{-8} k/(2 z_h^8)$) to account for the   $\rho$ large-$\omega^2$ dependence. 
At small  $T$ the spectral function displays  a  narrow resonance, with  vanishing width for $T\to 0$.  As $T$ increases changing the bulk geometry, the peak moves towards smaller values of $\omega^2$,  accompanied by a broadening of the line shape: the thermal effects on the gluonium  reduce the mass and make the state  unstable. At some value of $T$ the peak disappears from the spectral function,  indicating  the in-medium melting of the state.

\begin{figure}[t!]
\bec
\begin{tabular}{c c }
\includegraphics[width = 0.5\textwidth]{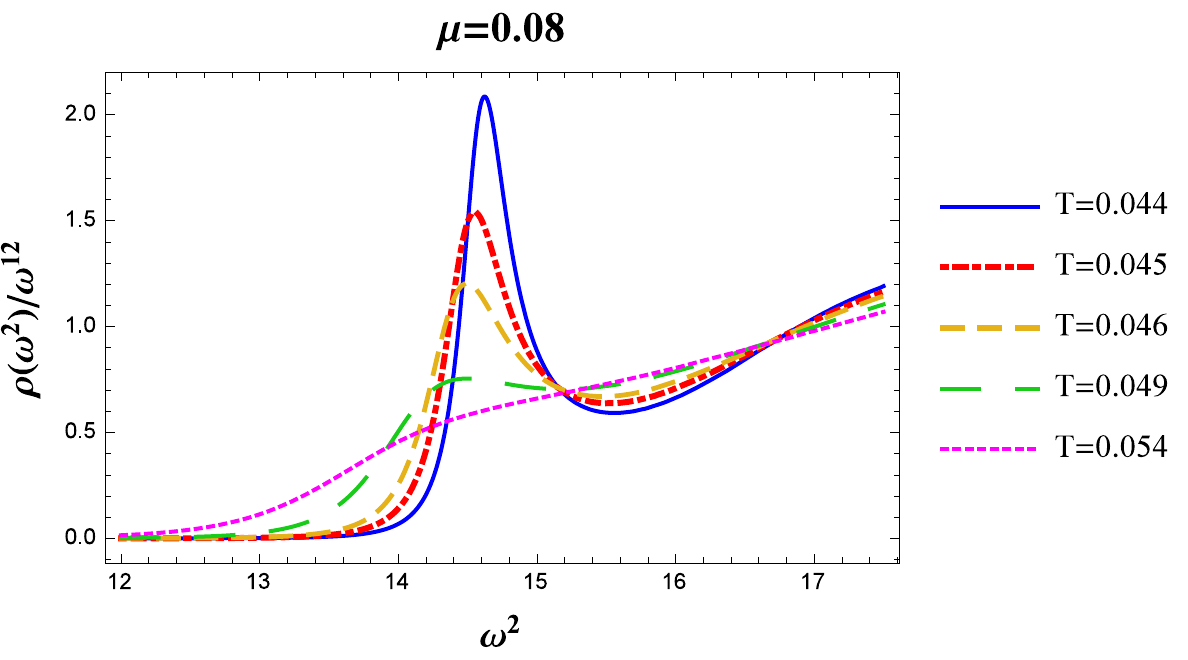}&
\includegraphics[width = 0.5\textwidth]{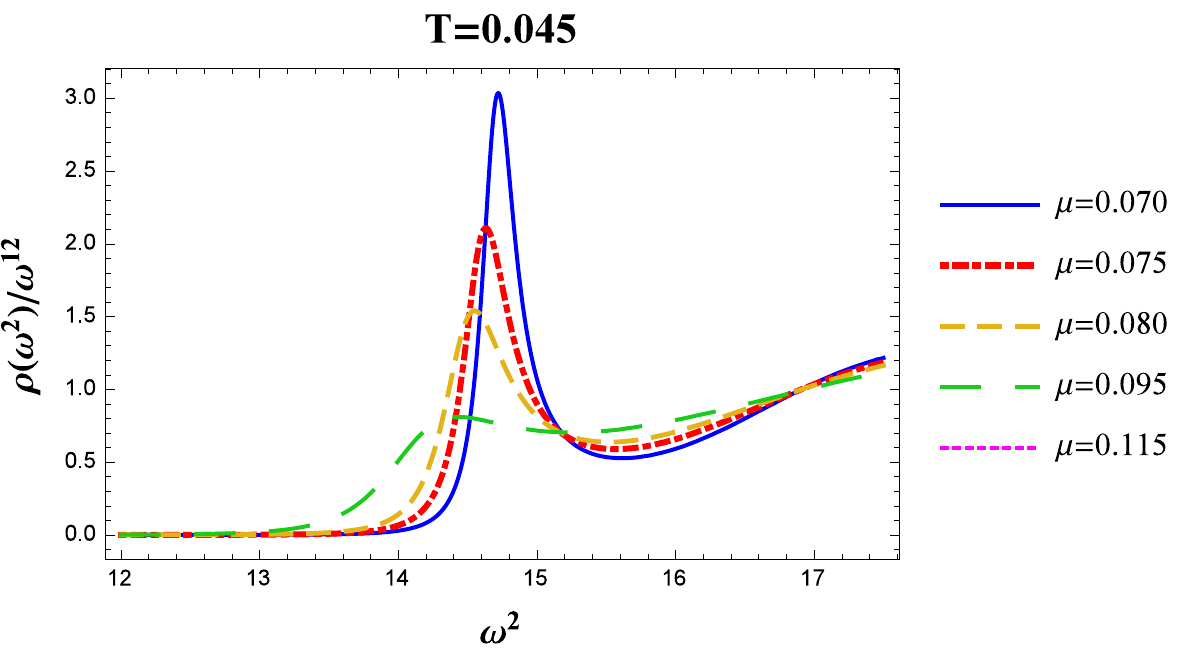}
\end{tabular}
\eec
\caption{\small  Spectral function $\rho(\omega^2)/(\omega^2)^6$ at fixed $\mu=0.08$  for different values of the temperature (left),  and  at fixed $T=0.045$  for different values of the chemical potential (right).
In the  plotted function   the constant  
$10^{-8} k/(2 z_h^8)$ has been factorized out. 
The dimensionful quantities are in units of $c$.}\label{rho-t-mu-1}
\end{figure}

Also at  finite  $\mu$  the peaks of the spectral function broaden and move  towards smaller values of $\omega^2$ as $T$ increases, up to a point  where no peak can be distinguished. 
The same behavior is  observed at fixed temperature, increasing the chemical potential. 
The broadening  is a signal that, as the temperature or  the quark chemical potential increases, the states become unstable,  getting a finite width (a quantum-mechanical argument for such a behavior is  in \cite{Colangelo:2012jy}). The results are collected in Fig.~\ref{rho-t-mu-1} for two values of $\mu$ and $T$.

\begin{figure}[t!]
\bec
\begin{tabular}{c c}
\includegraphics[width = 0.45\textwidth]{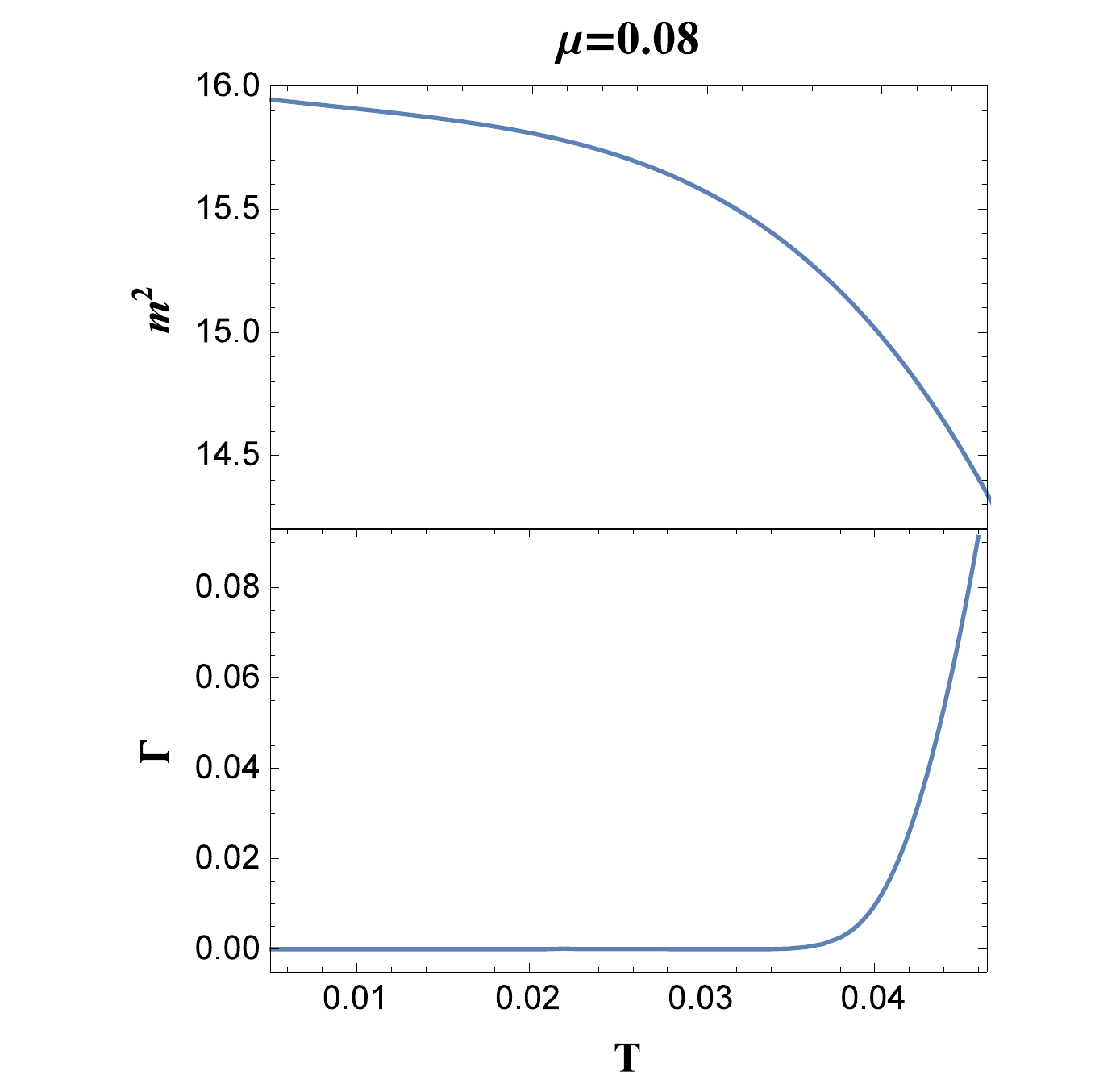}&
\includegraphics[width = 0.45\textwidth]{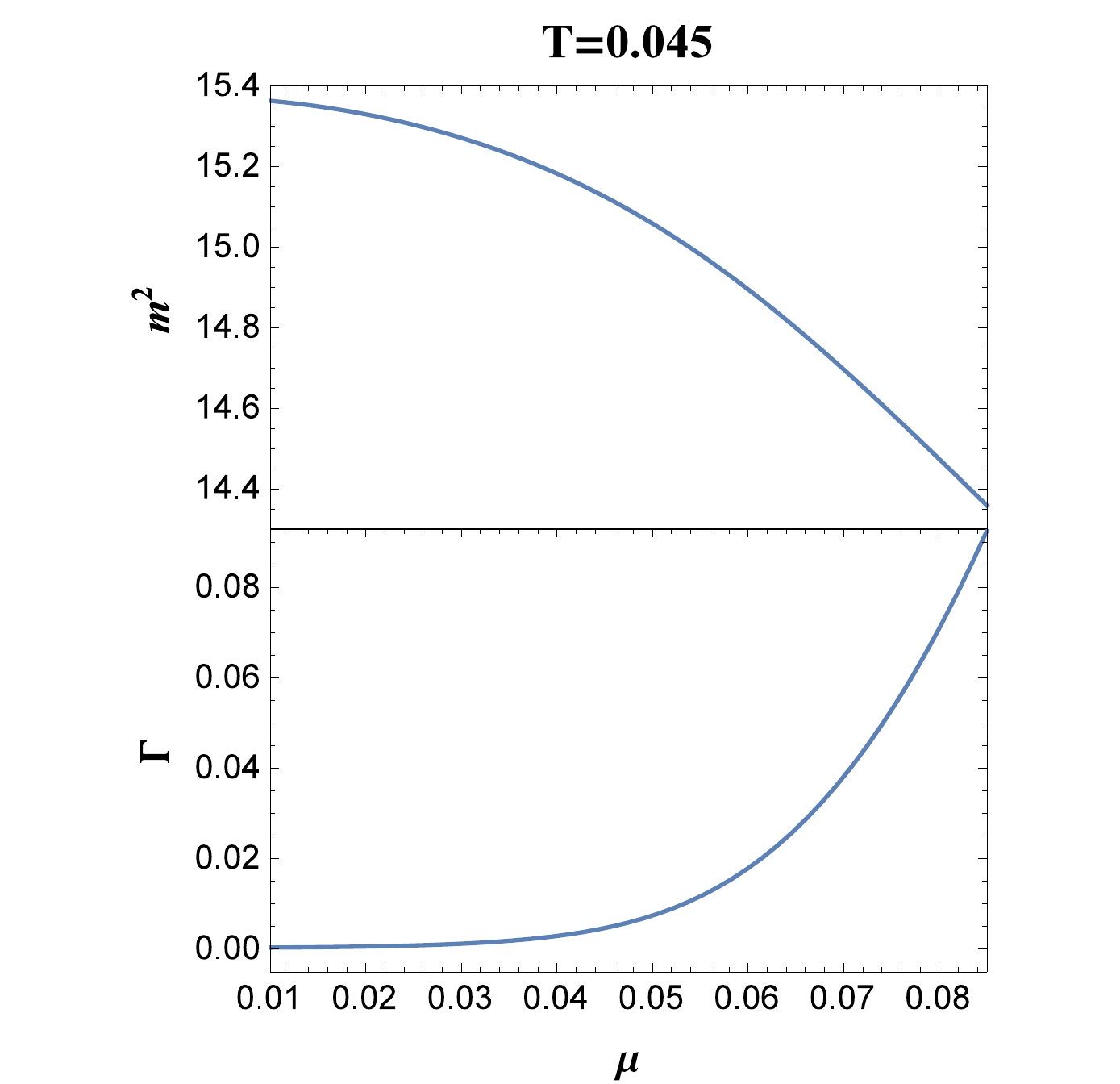}
\end{tabular}
\eec
\caption{\small  $m^2$ (up) and width (down) of the lightest $J^{PC}=0^{--}$ glueball,   at fixed $\mu=0.08$  for different values of temperature (left) and 
at fixed $T=0.045$  for different values of chemical potential (right), in the AdS/RN SW model. The dimensionful quantities are in units of $c$.}\label{m2-t}
\end{figure}

\begin{figure}[b!]
\begin{tabular}{cc}
\includegraphics[width = 0.45\textwidth]{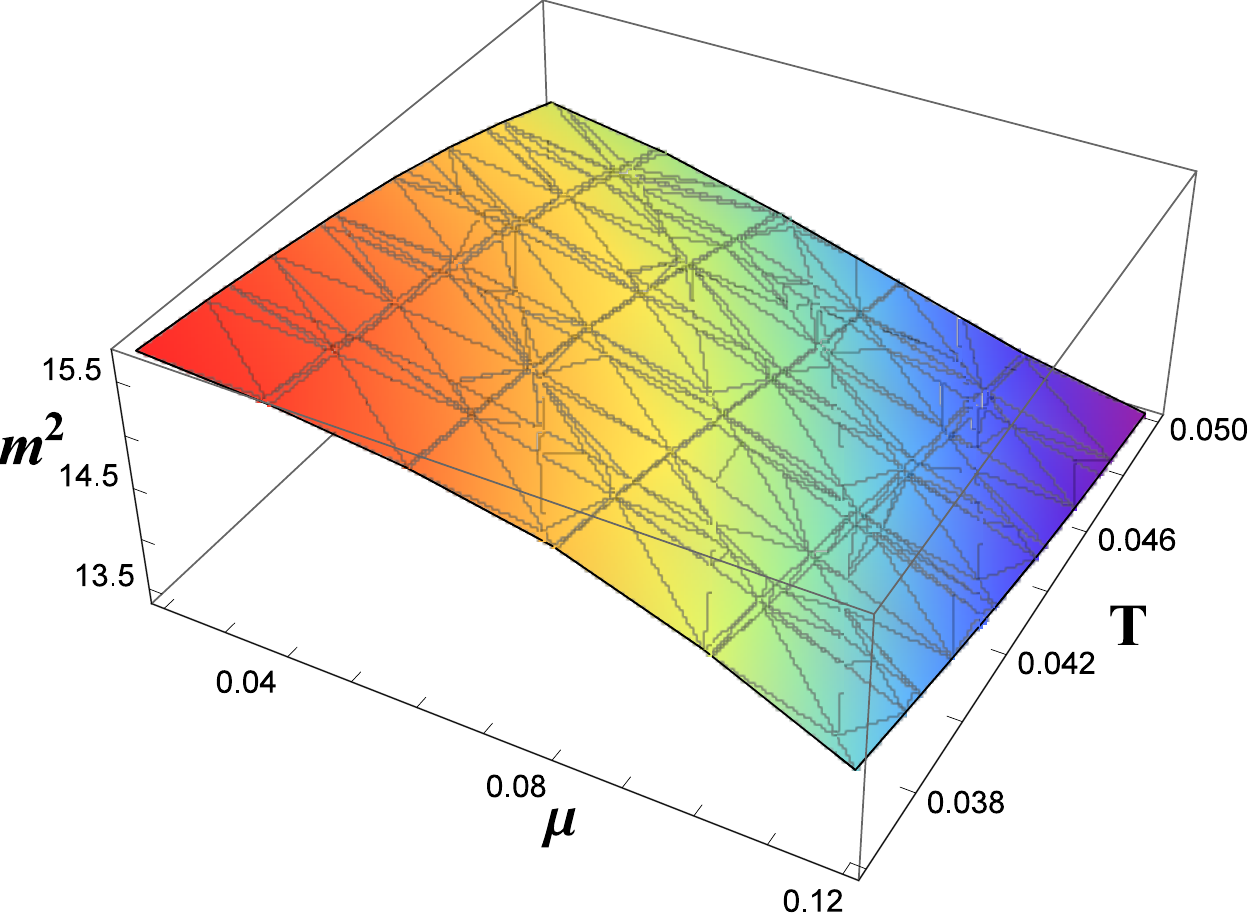}&
\includegraphics[width = 0.45\textwidth]{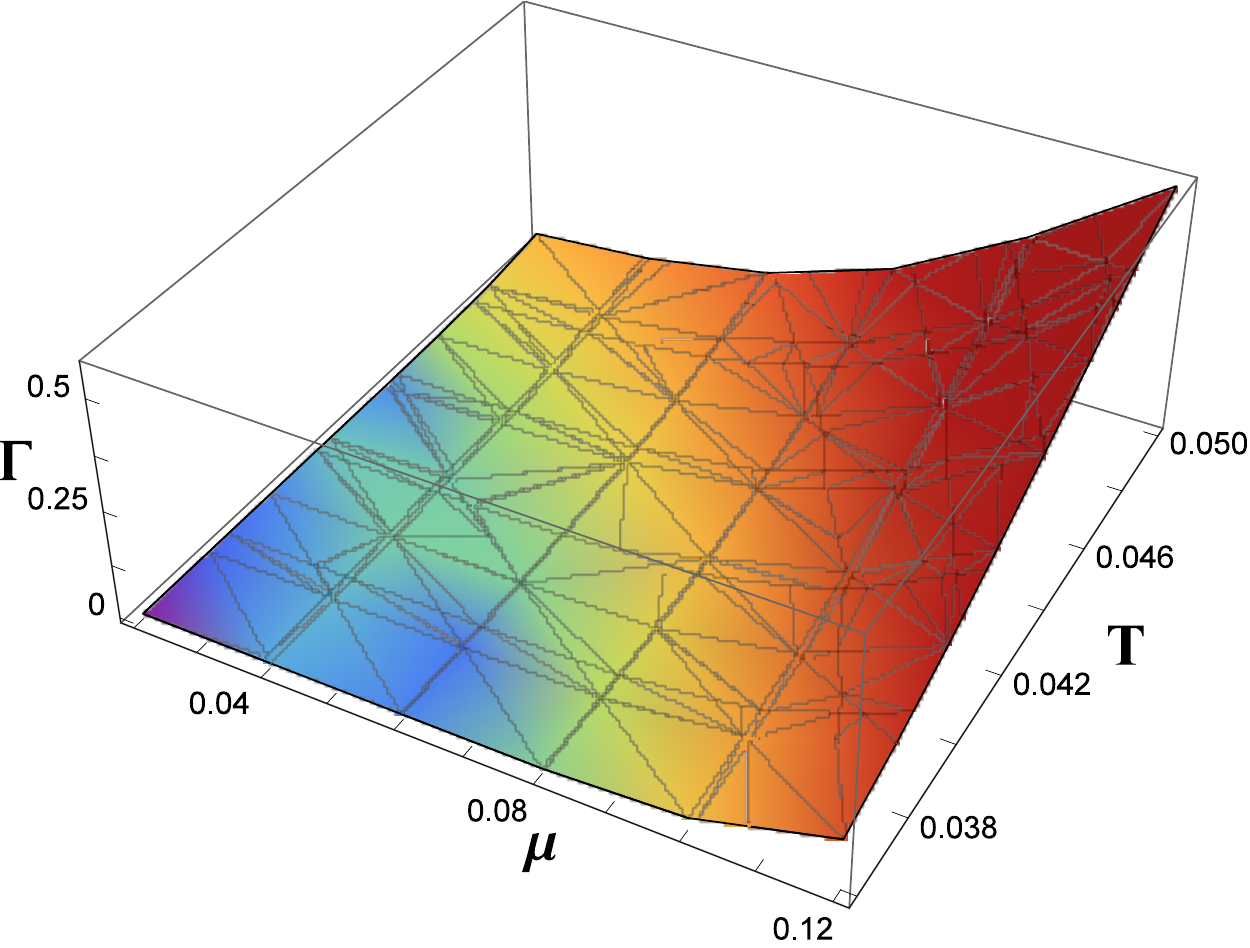}
\end{tabular}
\caption{\small   $m^2$ and width $\Gamma$  of the lightest $J^{PC}=0^{--}$ glueball, computed in a range of  temperature $T$ and  chemical potential $\mu$,  in the SW model with AdS/RN metric. The dimensionful quantities are in units of  the scale in the model, the parameter $c$.}\label{plot-3D}
\end{figure}

 To quantitatively extract the   temperature and chemical potential dependence of the lightest oddball mass, we  fit the peak in the spectral function  $\rho(\omega^2)$  using a Breit-Wigner-like expression \cite{Colangelo:2009ra,Fujita:2009wc}:
\begin{equation}\label{BWeq}
\rho_{BW}(\omega^2)=\frac{a\, (\omega^2)^b\, m\, \Gamma\, }{(\omega^2-m^2)^2+m^2\Gamma^2}\,,
\end{equation}
obtaining  the mass  $m$ and  width  $\Gamma$ of the state ($a$ and $b$ are parameters in the fit).
The melting temperature and  chemical potential are obtained 
looking at the values of $T, \mu$ where the peak in the spectral function is reduced
by a factor (we choose 20)  with respect to the point where the line-shape starts  broadening.
The results for two values of $T$ and $\mu$ are shown in Fig.~\ref{m2-t},
and a synopsis of the $T-\mu$ dependence is presented in  Fig.~\ref{plot-3D}.

 One can now make a comparison with other hadrons. Considering light vector and scalar $\bar q q $ mesons, the lightest scalar glueball and hybrid mesons, the values of $T$ and $\mu$  where the peak of the lowest lying oddball disappears from the spectral function are by far smaller  \cite{Colangelo:2012jy,Bellantuono:2014lra}.
This can be interpreted as  an indication of a higher sensitivity of this hadron to thermal and matter effects  at $T\neq 0$ and $\mu\neq 0$, and  that the  state is less stable than other conventional hadrons.

%

\vspace*{0.5cm}
 In the above discussion, we have assumed  the AdS/RN geometry as suitable  to formulate a QCD dual regardless of the values of $T$ and $\mu$. However, with this geometry it is known that duality holds above a line in the $T-\mu$ plane, where the deconfined phase is realized;  at small $T$ and $\mu$ AdS/RN  represents a metastable phase, the confining phase  being described, e.g. at $\mu=0$, by thermal-AdS. The transition between the two phases is holographically represented by a Hawking-Page transition \cite{Herzog:2006ra}. This can be seen considering in greater detail 
the limit $T \to 0$ with  finite chemical potential $\mu$.

In the AdS/Reissner-Nordstr\"om model the limit $T\to 0\,\,, \,  \mu\neq 0$ corresponds to  $Q^2 \to 2$  (while $z_h\to\infty$ corresponds to $T=0$ and $\mu=0$).
This is the case of an extremal black hole with coinciding  outer and inner horizon. The black hole function has a double zero in $z=z_h$:
\begin{equation}
 f(z) = \left(1-\frac{z}{z_h} \right)^2  \left(1+\frac{z}{z_h} \right)^2  \left(1+2 \frac{z^2}{z_h^2} \right) \,\, .
\end{equation}
Moreover,  at  $T=0$ the geometry has a horizon and  a  non-vanishing entropy, a  known feature of  models  based on the RN metric and used in the framework of the emergent quantum criticality.    Other consequences are in  the determination of  the spectral functions, where  the   behavior of the solution of the equation of motion near the horizon is needed.  For   $Q^2=2$  the asymptotic solution contains divergent terms proportional to $p_0^2$, which hinder the selection of the  in-falling condition. In studies of, e.g.,  transport coefficients at $T=0$ and  $\mu\neq 0$, the condition $\vec p\neq 0$  together with the limit $p_0 \to 0$  avoids divergences in the correlation functions \cite{Edalati:2009bi}. 

It has been proposed to study the points at  $T=0\,\,, \, \mu\neq 0$  considering a   model  having the function $f(z)$ in the  geometry  given by \cite{Lee:2009bya}
\begin{equation}\label{eq:tcAdS}
 f(z)=1+q^2 z^6\,
\end{equation}
and the time-component  $A_0(z)$ in \eqref{A0}. 
The metric   \eqref{RN-geom},\eqref{eq:tcAdS} is  solution of the Einstein equations, with the  condition $f(z_h)=0$  replaced by putting to zero the coefficient of $z^4$.  While in the HW model
the vanishing   $A_0(z_0)=0$ at the IR brane provides  a relation between $\mu$ and $q$,  in  the SW model,  a linear relation between $\mu$ and $q$  can be assumed, with the coefficient fixed  computing the boundary action \cite{Park:2011qq};   the  dilaton term in the action 
allows to cure the divergence of  $f$ and $A_0$  at large $z$ (naked singularity).  As in the thermal  AdS model,  the temperature can also be implemented using  a periodic Euclidean time coordinate.
The obtained geometry (thermal charged AdS - tcAdS)  is proposed as a dual of the confined phase of QCD at small temperature and  chemical potential, while the AdS/Reissner-Nordstr\"om model describes the deconfined phase, with a Hawking-Page transition between the two phases \cite{Lee:2009bya}. 

\begin{figure}[th!]
\bec
\includegraphics[width = 0.5\textwidth]{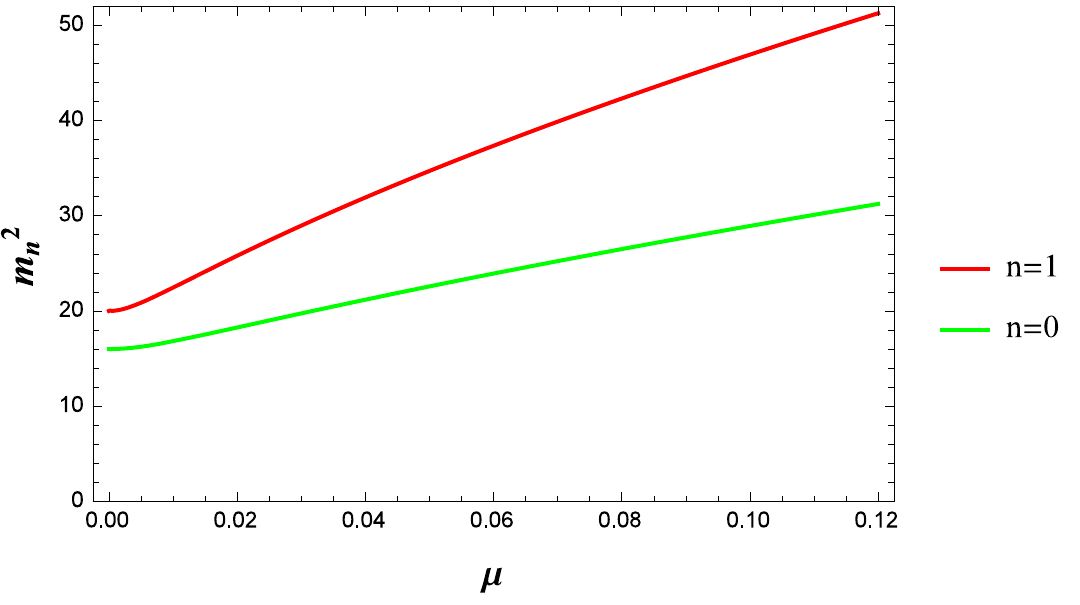}
\eec
\caption{\small   $m^2$ vs $\mu$ at $T=0$ for the first two states of the $J^{PC}=0^{--}$ gluonium,  in the SW   model with   geometry  \eqref{RN-geom},\eqref{eq:tcAdS}. The dimensionful quantities are in units of the scale $c$.}\label{sp-tcAdS}
\end{figure}

Using the  geometry  \eqref{RN-geom},\eqref{eq:tcAdS} at $T=0$, the two lightest gluonium states have mass   as depicted in Fig.~\ref{sp-tcAdS},  and at small $T$ the results  remain unaffected.  The   difference with respect to the AdS/RN model  is 
the increasing behaviour vs  $\mu$,   a confirmation  that the two models   describe  different phase  of QCD. 

The conclusion of the analysis is that, using AdS/RN, the $0^{--}$ oddball is more sensitive to matter effects than all other hadrons studied in the same framework, including the  $0^{++}$ glueball
and the $1^{+-}$ hybrid  \cite{Colangelo:2012jy,Bellantuono:2014lra}. On the other hand, with the tcAdS geometry a peculiar $\mu$-dependence of the hadron mass is found.

 \section{$J^{PC}=0^{--}$ glueball  phenomenology}\label{sec4}
In our calculations the lowest gluonium state with  $J^{PC}=0^{--}$  is quite light. On the basis of this result,  we  select suitable processes  for the production and the identification of this state.~\footnote{A few modes are mentioned  in   Ref.~\cite{Qiao:2014vva}.}
Our guidelines are the quantum number selection rules,  since the relevant hadronic couplings cannot be  computed  in  the  models we are using here.
For definiteness, we  consider the lightest oddball with mass $m_{0^{--}}=2.8$ GeV.

Production modes in radiative and hadronic two-body transitions, occurring in different waves, are collected in Table~\ref{tab1}.
Spin $1^{++}$ charmonium and bottomonium states decay radiatively in the $0^{--}$ oddball in S- and D-wave, while $2^{++}$ states decay in D-wave.
The suppression  $ \frac{\Gamma(\chi_{c1}(1P) \to \gamma \, G(0^{--}))}{\Gamma(J/\psi \to \gamma \, G(0^{++}))} \simeq \alpha_s(m_c)$ is expected, while
 radiative processes in bottomonium,    which are phase-space favored,  are  suppressed by  $(e_b/e_c)^2$ with respect to the corresponding charmonium rates.
The  hadronic decay mode $X(3872)\to \omega \, G(0^{--})$ is at the limit of the phase space, and  the bottomonium modes $\chi_{b1}(nP) \to \omega \, G(0^{--})$  are allowed. Another interesting process involves the isoscalar scalar $\pi \pi$ configuration in the final state, namely  $h_c(1P) \to \pi^+ \pi^- G(0^{--})$, together with the bottomonium counterpart, the 
 $h_b(1P) \to f_0(980) G(0^{--})$ and  $h_b(2P) \to f_0(980) G(0^{--})$  transitions.
Other charmonium decays in P-wave, namely $\psi(nS) \to G(0^{--}) \chi_{c0}(1P)$ and $\chi_{c1}(nP) \to G(0^{--}) J/\psi$ (with the corresponding   bottomonium transitions),  are only possible for very heavy $\bar c  c$ ($\bar b  b$) decaying states.
We remark 
the  P-wave  two-body  $h_b(nP)$ decays in  $J^{PC}=0^{++}$ scalar glueball and  $J^{PC}=0^{--}$ oddball, which are very peculiar modes
for the exclusive gluonium production.

\begin{table}[t]
\bec
\begin{tabular}{| c | c |}
\hline
 radiative transition&  \\
\hline 
$\chi_{c1}(3510) \to \gamma \, G(0^{--})$&$\chi_{b1}(9892) \to \gamma \, G(0^{--})$\\
$X(3872) \to \gamma \, G(0^{--})$&$\chi_{b1}(10255) \to \gamma \, G(0^{--})$\\ & \\
$\chi_{c2}(3556) \to \gamma \, G(0^{--})$&$\chi_{b2}(9912) \to \gamma \, G(0^{--})$\\
$\chi_{c2}(3927) \to \gamma \, G(0^{--})$&$\chi_{b2}(10269) \to \gamma \, G(0^{--})$ \\ 
\hline 
 hadronic transition &  \\ \hline
$X(3872) \to \omega \, G(0^{--})$&$\chi_{b1}(10255) \to (\omega, \phi, J/\Psi)  \, G(0^{--})$\\
&$\Upsilon(nS) \to (f_1(1270), \chi_{c1}, X(3872))  \, G(0^{--})$\\ &\\
$h_{c}(3525) \to \pi \pi \, (I=0)  \, G(0^{--})$&$h_{b}(9899) \to f_0(980) \, G(0^{--})$\\
&$h_{b}(10260) \to f_0(980) \, G(0^{--})$\\ 
&$h_{b}(9899) \to G(0^{++}) \, G(0^{--})$\\ 
&$h_{b}(10260) \to G(0^{++}) \, G(0^{--})$\\
 \hline
\end{tabular}
\caption{Production modes of the $J^{PC}=0^{--}$ glueball, for $m_{0^{--}}=2.8$ GeV.}\label{tab1}
\eec
\end{table}

 Decay modes  of the $J^{PC}=0^{--}$  glueball are listed in Table~\ref{tab2}. In addition to the modes involving the axial $I=0$ $f_1(1270)$ meson, it is worth mentioning the full set of $P-$wave decays, among which
there is $\rho \pi (I=0)$. 
The couplings governing the various modes  cannot be computed in the framework discussed here,  and require specific calculations  deferred  to a dedicated study. 

\begin{table}[t]
\bec
\begin{tabular}{| l |  }
\hline
mode\\
\hline 
$ G(0^{--}) \to \gamma \, f_1(1270)$\\
$ G(0^{--}) \to \omega \, f_1(1270)$\\
$ G(0^{--}) \to \rho \, a_1(1260)\,\,\,(I=0)$\\ \\
$ G(0^{--}) \to h_1(1270) \, f_0(980)$\\ 
$ G(0^{--}) \to \rho \, \pi \,\,\,(I=0) $\\
$ G(0^{--}) \to K^* \, K \,\,\,(I=0) $\\
$ G(0^{--}) \to (\eta,\eta^\prime)(\omega, \phi) $\\
 \hline
\end{tabular}
\caption{Decay modes of the $J^{PC}=0^{--}$ glueball, $m_{0^{--}}=2.8$ GeV. }\label{tab2}
\eec
\end{table}

\section{Conclusions}\label{conc}

Our main result is that the lowest-lying $J^{PC}=0^{--}$ glueball, examined in different bottom-up holographic models of QCD, is lighter than envisaged by other  approaches. This opens interesting possibilities for the experimental search of this unconventional hadron. We have also investigated the in-medium effects,
 obtaining that using AdS/RN the $0^{--}$ oddball is more sensitive to matter effects than all other hadrons studied in the same framework. On the other hand, with the tcAdS geometry a peculiar $\mu$-dependence of the hadron mass is found.   Several production and decay modes can be exploited  for the search of this elusive gluonium resonance.

As a final remark, we find  inspiring that the lowest mass we have obtained using the SW model is  close to the $D^0$ mass, in view of the dominance observed in  
$D^0 \to \pi^+ \pi^- \pi^0$  of an  exotic $J^{PC}=0^{--}$ isoscalar state. It is worth reconsidering this issue in a dedicated study.

\vspace*{1cm}
\noindent {\bf Acknowledgments.}\\
\noindent We thank Fulvia De Fazio and Stefano Nicotri  for  discussions.

\bibliographystyle{JHEP}
\bibliography{our-ref}
\end{document}
\